\documentclass[aps,twocolumn,showpacs]{revtex4}
\usepackage{graphics}
\begin{document}
\bibliographystyle{apsrev}
\def\half{{1\over 2}}
\def \D {\mbox{D}}
\def\curl {\mbox{curl}\,}
\def \ep {\varepsilon}
\def \lleq {\lower0.9ex\hbox{ $\buildrel < \over \sim$} ~}
\def \ggeq {\lower0.9ex\hbox{ $\buildrel > \over \sim$} ~}
\def\beq{\begin{equation}}
\def\eeq{\end{equation}}
\def\ber{\begin{eqnarray}}
\def\eer{\end{eqnarray}}
\def \apl {ApJ, }
\def \aps {ApJS, }
\def \pd {Phys. Rev. D, }
\def \prl {Phys. Rev. Lett., }
\def \pl {Phys. Lett., }
\def \np {Nucl. Phys., }
\def \l {\Lambda}

\def\apj{{Astroph.\@ J.\ }}
\def\mn{{Mon.\@ Not.\@ Roy.\@ Ast.\@ Soc.\ }}
\def\asta{{Astron.\@ Astrophys.\ }}
\def\aj{{Astron.\@ J.\ }}
\def\prl{{Phys.\@ Rev.\@ Lett.\ }}
\def\pd{{Phys.\@ Rev.\@ D\ }}
\def\nucp{{Nucl.\@ Phys.\ }}
\def\nat{{Nature\ }}
\def\plb {{Phys.\@ Lett.\@ B\ }}
\def \jetpl {JETP Lett.\ }

\title{Quintessential Inflation on the Brane \\and the\\ Relic 
Gravity Wave Background}
\author{M. Sami}
\altaffiliation[On leave from:]{ Department of Physics, Jamia Millia, New Delhi-110025}
\email{sami@iucaa.ernet.in}
\author{V. Sahni}
\email{varun@iucaa.ernet.in}
\affiliation{IUCAA, Post Bag 4, Ganeshkhind,\\
 Pune 411 007, India.}  
 
 
\begin{abstract}
Quintessential inflation describes a scenario in which both inflation
and dark energy (quintessence) are described by
the same scalar field. In conventional braneworld models of
quintessential inflation gravitational particle production
is used to reheat the universe. This  reheating mechanism is very inefficient
and results in an excessive
production of gravity waves which 
violate nucleosynthesis constraints and invalidate the model.
We describe a new method of realizing quintessential inflation on the brane
in which inflation is followed by `instant preheating' (Felder, Kofman \& Linde 1999).
The larger reheating temperature in this model results in a 
smaller amplitude of relic gravity waves which is consistent with nucleosynthesis
bounds.
The relic gravity wave background has a `blue' spectrum at high frequencies
and is a generic byproduct of successful quintessential inflation on the brane.
\end{abstract} 

\maketitle

\section{Introduction}

One of the most remarkable discoveries of the past decade is that the universe is
accelerating. An accelerating universe is supported by observations of high
redshift type Ia supernovae treated as standardized candles \cite{sn1,sn2} 
and, more indirectly,
by observations of the cosmic microwave background and galaxy clustering
\citep{wmap,tegmark03}.
Within the framework of general relativity, cosmic acceleration should be
sourced by an energy-momentum tensor which has a large negative 
pressure (dark energy). The simplest form of dark energy is undoubtedly the
cosmological constant for which $p = -\rho$ = constant. However, due to its
un-evolving nature, the cosmological constant must be set to an extremely small
value in order to dominate the expansion dynamics of the universe at precisely
the present epoch. This gives rise (according to one's viewpoint) either to a
fine-tuning problem or to a `cosmic coincidence' problem. For this reason theorists
have suggested making dark energy a dynamical quantity associated
with a Lagrangian and having well defined equations of motion (see 
\cite{ss00,carroll01,pr02,sahni02a,paddy03} for reviews
of dark energy). Perhaps the simplest dynamically evolving dark energy
models are quintessence fields -- scalar fields which couple minimally
to gravity and which roll down a steep potential \cite{peebles88,wetterich88}.
Although quintessence models do not resolve the `cosmic coincidence'
conundrum they do alleviate, to some extent, the fine-tuning problem faced by 
the cosmological constant
since they approach a common evolutionary path from a wide range of initial conditions.

Braneworld models \cite{randall,shiromizu} add an interesting new dimension to 
scalar field dynamics on the brane.
The presence of a quadratic density term (due to high energy corrections) 
in the Friedman equation on the brane fundamentally alters the 
expansion dynamics at early epochs by greatly increasing the Hubble parameter
and hence the damping experienced by the scalar field as it rolls down its potential
\cite{maartens}.
Consequently, 
inflation on the brane can be realized by very steep potentials 
-- precisely those used to describe quintessence.
The braneworld scenario therefore provides us with the opportunity 
to unify inflation  and
dark energy through a mechanism called 
quintessential inflation \cite{pv99,copeland,lidsey1,sss02,majumdar,nkd}. 
Models of quintessential inflation have a single major
drawback: they are usually derived from {\em non-oscillating} potentials
for which the standard reheating mechanism does not work.
Indeed, in order to ensure that the inflaton survives
until today one usually
invokes a method of reheating based on the quantum mechanical production of 
particles in the time-varying gravitational field after inflation \cite{ford,spokoiny}.
This method of reheating is very inefficient, and leads to a 
`kinetic regime' of prolonged duration
when braneworld corrections are no longer
important and the scalar field rapidly drops down a steep potential, resulting
in $p_\phi \simeq \rho_\phi \simeq {\dot\phi^2}/2$ and $a \propto t^{1/3}$. 
Gravity waves,
created quantum mechanically during the kinetic regime have a `blue tilt'
and, for a prolonged kinetic regime,
their energy density can dominate the energy density of the universe
and violate nucleosynthesis constraints \cite{sss02}
(see also \cite{star79,sahni90,ss92,giovan98,lmw00,kobayashi,hiramatsu,elmw03}).
Thus conventional braneworld 
models of quintessential inflation run into serious problems 
associated with copious graviton production which renders them unviable for an 
extended region in parameter space.
As we shall show in this paper, this problem is easily circumvented if, instead of
gravitational particle production, one invokes an alternative method of reheating,
namely `instant preheating' proposed by Felder, Kofman and Linde 
\cite{FKL,FKL1,kls97} (see also
\cite{kls94,stb95,R,chiba}).
This method results in a much higher reheat temperature and therefore in a much
shorter duration kinetic regime. As a result the amplitude of relic 
gravity waves
is greatly reduced and there is no longer any conflict with nucleosynthesis
constraints. 
(For other approaches to reheating in quintessential inflation see
\cite{curvaton1,dimo,curvaton2,bi1,bi2}. An alternate approach to dark energy in braneworld models
is provided in \cite{DDG,ss03}.)
Before we apply the instant preheating method to quintessential
inflation on the brane let us briefly review scalar field dynamics 
in braneworld cosmology.

\section{Braneworld Inflation with an Exponential Potential}

It is well known that braneworld models can give rise to successful inflation
by greatly expanding the parameter space for which inflation can take place
\cite{copeland,lidsey1,sss02}.
This is true both for the braneworld based on the Randall-Sundrum II scenario,
(henceforth RS II)
as well as in Gauss-Bonnet cosmology, for which the five dimensional action is
(see \cite{lidsey03,Nojiri} and references therein)
\ber
S &=& \frac{M_5^3}{16\pi}\int d^5x\sqrt{-g}\big\lbrace {\cal R} - 2\Lambda_5
+ \alpha_{GB} \lbrack {\cal R}^2 - 4{\cal R}_{AB}{\cal R}^{AB} \nonumber\\ 
&+& {\cal R}_{ABCD} {\cal R}^{ABCD}\rbrack \big \rbrace 
+ \int d^4x\sqrt{-h} ({\cal L}_m - \lambda)~,
\label{action}
\eer
${\cal R}, R$ refer to the Ricci scalars in the metric $g_{AB}$ and 
$h_{AB}$; $\alpha_{GB}$ has dimensions of ({\em length})$^2$ and
is the Gauss-Bonnet coupling, while $\lambda$ is the brane tension.
The analysis of cosmological dynamics based upon action~(\ref{action})
, shows that there is a characteristic GB energy
scale $M_{\rm GB}$\cite{DR} such that,
 \begin{eqnarray}
\rho\gg M_{\rm GB}^4~& \Rightarrow &~ H^2\approx \left[ {\kappa_5^2 \over
16\alpha_{\rm GB}}\, \rho \right]^{2/3}~(GB)\,,\label{vhe}\\
M_{\rm GB}^4 \gg \rho\gg\lambda~& \Rightarrow &~ H^2\approx {\kappa_4^2
\over
6\lambda}\, \rho^{2}~~~~~~~~~~~~~(RS)\,,\label{he}\\
\rho\ll\lambda~& \Rightarrow &~ H^2\approx {\kappa_4^2 \over 3}\,
\rho~~~~~~~~~~~~~~(GR)\,. \label{gr}
 \end{eqnarray}

The modified  Einstein equations on the brane contain 
high-energy corrections as well as the projection of the Weyl tensor from the 
bulk on to the brane. The Friedmann equation
on the RS brane is
\begin{equation}
H^2={1 \over 3 M_p^2} \rho \left(1+{\rho \over {2\lambda}}\right)~, ~M_p = \frac{M_4}{\sqrt{8\pi}}~,
\label{friedman}
\end{equation}
(we have dropped the {\it dark radiation} term and the effective four dimensional
cosmological constant since neither is likely to be 
relevant for inflation).  
Our discussion below is quite general,
and can easily incorporate the effect of GB correction.

A homogeneous scalar field which couples minimally gravity and propagates on the brane
satisfies the evolution equation
\begin{equation}
\ddot{\phi}(t)+3H\dot{\phi}(t)+V_{,\phi}=0~,
\end{equation}
and its energy density and pressure are given by
\begin{equation}
\rho_{\phi}={ \dot{\phi}^2 \over 2}+V(\phi),~~~~p_{\phi}={ \dot{\phi}^2 \over 2}-V(\phi)~.
\end{equation}  

We shall assume that for $|\phi|//M_p \gg 1$ the scalar-field
potential can be described by a pure exponential
\begin{equation}
V(\phi)=V_0e^{{\alpha\phi}/M_p}~.
\label{exp}
\end{equation}
For $\alpha > \sqrt{2}$, the exponential potential (\ref{exp}) is too 
steep to sustain inflation in standard cosmology.
However, as discussed in the introduction, such
steep potentials can easily drive inflation in braneworld cosmology  \cite{copeland}, 
\footnote{see \cite{gman} for a different approach to steep inflation.}.
In this case,
the increased damping
due to the quadratic term in (\ref{friedman})
leads to slow-roll inflation on the brane at high energies, when $\rho/\lambda \gg 1$.
In this limit, the slow-roll equations become \cite{copeland,sss02} 
\ber
{\dot{a}(t) \over a(t)}&=&\sqrt{{1 \over {6M_p^2 \lambda}}} V(\phi)~,\nonumber\\
3H\dot{\phi}(t)&=& -V_{,\phi}~,
\label{s1}
\eer
so that
\begin{equation}
\dot{\phi}=-\alpha \sqrt{{2 \lambda} \over 3}~.
\label{velo}
\end{equation} 
As a result
\beq
\phi(t)= \phi_i - \sqrt{\frac{2\lambda_b}{3}} \alpha (t-t_i).
\label{eq:motion}
\eeq
The scale factor obtained from (\ref{s1}) passes through an inflection point 
marking the end of inflation, the corresponding scalar field value is \cite{sss02}
\beq
\phi_{\rm end} =
-\frac{M_P}{\alpha}
\log{\bigg (\frac{V_0}{2\lambda\alpha^2}\bigg )},
\label{phiend}
\eeq
substitution in (\ref{exp}) leads to 
\beq
V_{\rm end} \equiv V_0e^{\alpha\phi_{\rm end}/M_P} =
2\lambda\alpha^2\\
\label{eq:Vend}.
\eeq
After inflation ends, it takes a little while for brane corrections to disappear 
and for the kinetic regime to commence.
The passage from the end of inflation to the commencement
of the kinetic regime is reflected in the following fitting formula
which relates the Hubble parameter at the onset of the kinetic regime to its counterpart at the end of inflation \cite{sss02}
\begin{equation}
H_{kin} = H_{end} F(\alpha),~~~(\alpha \ggeq 3) 
\label{Hkin}
\end{equation}
where the fitting formula $F(\alpha)=0.085 -{0.688 \over \alpha^2}$
was obtained by numerically integrating the equations of motion.
A brief comment is in order on the influence of brane term, $ B_{brane}$ after inflation
($ H^2 \sim \rho B_{brane}$). Around $\phi=\phi_{end}$, we can approximately write $ B_{brane}$ as\cite{sss02},
$ B_{brane} \simeq 1+\alpha^2 e^{-\alpha(\phi_{end}-\phi)/M_p}$. Hence the influence of brane term diminishes faster for larger values of $\alpha$ which manifests in the fitting formulas (\ref{Hkin}) and others which occur
in text below.\\
 
The COBE normalized amplitude for density perturbations allows us
to determine the value of the potential at the end of inflation
and the value of brane tension
\ber
V_{end} &\simeq& {{3 \times 10^{-7}} \over \alpha^4} 
\left({M_p \over {\cal N }+1}\right)^4\frac{1}{(1 + A_T^2/A_s^2)^{1/4}}
~,\nonumber\\
\lambda &\simeq& {{1.3 \times 10^{-7}} \over \alpha^6} 
\left({M_p \over {\cal N }+1}\right)^4\frac{1}{(1 + A_T^2/A_s^2)^{1/4}}   ~,
\label{vend}
\eer
from which one fixes the value of $\dot{\phi}$ at the end of inflation
\begin{equation}
\dot{\phi}_{end}=-\sqrt{V_{end} \over 3}~,
\label{PHIDOT} 
\end{equation} 
here $A_T$ and $A_S$ are the amplitudes of tensor and scalar modes
created during inflation \cite{maartens,lmw00},
and $ {\cal N }$ is the number of {\it e}-foldings which we take to be
$ {\cal N }\simeq 70$. 
At the end of inflation, in addition to the scalar field energy density, a small amount
of radiation is also present $\rho_r \simeq 0.01 g_p H_{end}^4$,
arising due to particles produced quantum 
mechanically by the changing gravitational field after inflation 
($g_p$ is the number of particle species created) \cite{ford,spokoiny}.
Using equations (\ref{friedman}) and (\ref{vend})
one finds $\rho_{\phi}/\rho_r \sim  10^{16}g_p^{-1}$. 
This implies that equality between the inflaton field and radiation 
($\rho_{\phi} \simeq \rho_r$) is reached very late when the temperature has 
dropped below \cite{sss02}
\beq
T_{eq} \lleq \sqrt{g_p} {\rm GeV}~,
\label{temp}
\eeq
for $\alpha \ggeq 5$.
If gravitational particle production is the sole means of generating reheating, then,
prior to the radiative regime ($T > T_{eq}$)
the universe will enter into an extended kinetic
regime, during which $\rho_\phi \gg \rho_r$ and $p_\phi \simeq \rho_\phi$ so that
$a(t) \propto t^{1/3}$. 

It is well known that an inflationary universe gives rise to a stochastic background
of relic gravity waves of quantum mechanical origin  \cite{star79}. 
If the post-inflationary epoch is 
characterised by an equation of state $w = p/\rho$ then the spectral energy
density of gravity waves produced during slow-roll inflation is \cite{sahni90}
\beq
\rho_g(k) \propto k^{2\left (\frac{w - 1/3}{w + 1/3}\right )}~.
\eeq
In our braneworld model 
$w \simeq 1$ during the kinetic regime, consequently the gravity wave background 
generated during this epoch will have a 
blue spectrum $\rho_g(k) \propto k$. 
In this case it can be shown that the energy density of relic gravity waves 
at the commencement of the radiative regime is given by 
\cite{sss02}
\begin{equation}
\left({\rho_g \over \rho_{r}} \right)_{eq}={64 \over {3 \pi}} h_{GW}^2 \left({T_{kin} \over T_{eq}} \right)^2~,
\label{graden}
\end{equation}
where $h_{GW}$ is the dimensionless amplitude of gravity waves and
COBE normalization (with ${\cal N} \simeq 70$) gives
\begin{equation}
h_{GW}^2 \simeq 1.7\times 10^{-10}~.
\end{equation} 
Substituting for $T_{kin}$ from (\ref{Hkin}) and (\ref{vend}) and 
for $T_{eq}$ from (\ref{temp}) we find that
$\left(\rho_g /\rho_{r} \right)_{eq} \gg 1$ in this scenario.
Thus braneworld inflation with a steep potential invariably results in an
over-production of gravity waves which grossly violate the nucleosynthesis
bound $\left(\rho_g/ \rho_{r} \right)_{eq} \lleq 0.2$.

Before showing how instant preheating can resolve this issue let us mention
a few important formulae which will help in clarifying the issues involved.
As mentioned earlier, the commencement of the kinetic regime is not instantaneous
and brane effects persist for some time after inflation has ended. The 
temperature at the commencement of the kinetic regime $T_{kin}$ is related to the 
temperature at the end of inflation by
\begin{equation}
T_{kin}=T_{end}\left(a_{end} \over a_{kin} \right) = T_{end} F_1(\alpha)
\label{Tkin}
\end{equation}
where  $F_1(\alpha)=\left(c+{d\over \alpha^2} \right)$, $c \simeq 0.142$,~ $d \simeq -1.057$ and 
$T_{end} \simeq \left(\rho_{r}^{end} \right)^{1/4}$. The equality between the 
scalar field and radiation densities  takes place at the temperature
\begin{equation}
T_{eq}= T_{end} {F_2(\alpha) \over {\left(\rho_{\phi}/\rho_{r}\right)_{end}^{1/2}}}
\label{Teq}~,
\end{equation}
where
$ F_2(\alpha)=\left(e+{f \over \alpha^2} \right)$, $e\simeq 0.0265$,~$f\simeq -0.176$. 
(The fitting formulas (\ref{Tkin}) and (\ref{Teq}) were obtained 
in \cite{sss02} by numerically integrating the equations of motion.)
A useful formula relates
the Hubble parameters at the commencement of the radiative and kinetic regimes
\begin{equation}
H_{eq}=\sqrt{2} H_{kin} \left({T_{eq} \over T_{kin}} \right)^3~.
\label{Hrad}
\end{equation} 

Using equations (\ref{Tkin}), (\ref{Teq}) and (\ref{graden}) we obtain 
a useful expression which links the ratio of the scalar-field density to the 
radiation density at the end of inflation
with the ratio between the gravity wave density and the radiation density at the start
of the radiative era
\begin{equation}
\left({\rho_{\phi} \over \rho_{r}}\right)_{end}=
{3 \pi \over  64 }\left({F_2^2(\alpha) \over {h_{GW}^2 F_1^2(\alpha)}}\right) \left({\rho_g \over \rho_{r}} \right)_{eq}~.
\label{endratio}
\end{equation}  
Equation (\ref{endratio}) is an important result since it allows us to
set a limit to the ratio between the scalar field density and the 
radiation 
density at the end of inflation. 
Nucleosynthesis constraints provide an upper bound on the energy density in gravity
waves at the start of the radiative era:
$(\rho_g/\rho_{r})_{eq} \lleq 0.2$; substituting
in (\ref{endratio}), and assuming a steep potential 
($\alpha\ggeq 5$), we arrive at the following result
\begin{equation}
\left(\rho_{\phi}/\rho_{r}\right)_{end} \lleq 10^7~.
\label{endratio1}
\end{equation} 
Since $\left(\rho_{\phi}/\rho_{r}\right)_{end} \sim 10^{16}g_p^{-1}$ 
in the case of reheating sourced by
gravitational particle production \cite{sss02}, this mechanism leads to the
nucleosynthesis constraint being violated by almost nine orders of magnitude 
unless $g_p \ggeq 10^9$ ! ($g_p$ is the number of particle species created using
this mechanism.) 
Next, we shall examine the instant preheating mechanism
discovered by Felder, Kofman and Linde (1999) and show that it is capable of producing 
$\left(\rho_{\phi}/\rho_{r}\right)_{end}$ consistent with (\ref{endratio}) and (\ref{endratio1}). 

\section{Braneworld Inflation Followed by Instant Preheating}

Braneworld Inflation induced by the steep exponential potential (\ref{exp}) ends when
$\phi=\phi_{end}$, see (\ref{phiend}). 
Without loss of generality, we can make the inflation end at the
origin by translating the field
\begin{equation}
V(\phi') \equiv V(\phi)=\tilde{ V_0} e^{\alpha \phi'/M_p}~,
\label{texp}
\end{equation}
where $\tilde{ V_0}=V_0e^{+\alpha \phi_{end}}/M_p $ and $\phi'=\phi-\phi_{end}$.
In order to achieve reheating after inflation has ended we assume that the inflaton 
$\phi$ interacts with
another scalar field $\chi$ which has a Yukawa-type interaction with 
a Fermi field $\psi$. The interaction Lagrangian is

\begin{figure}
\resizebox{3.0in}{!}{\includegraphics{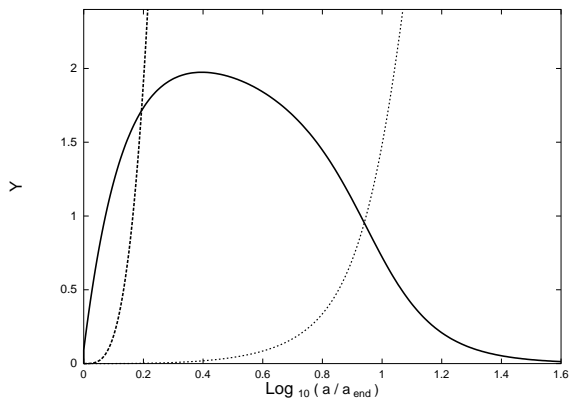}}
\caption{ The post inflationary evolution of $Y=10^9 \times |\dot{\phi}|/M_p^2 $,~ $g \phi^2/M_p^2$ is shown as a function of the scale factor for the model described by
Eq (\ref{cosine}) with $V_0\simeq 10^{-30} M_p$,
$\tilde{\alpha}=50$ and $p=0.1$. 
The solid curve corresponds to $\dot{\phi}/M_p^2$ whereas the dashed and dotted curves 
correspond to
$ g \phi^2/M_p^2$ for $g=10^{-6}$ (extreme left) and $g=10^{-9}$ (extreme right) 
respectively. The violation of the adiabaticity condition (necessary
for particle production to take place) arises in the region bounded by the solid and dashed curves. The time duration of
particle production is seen to be smaller for larger values of the coupling $g$. 
For the range of $g$
allowed by
the nucleosynthesis constraint, particle production is almost instantaneous.}
\label{nadiabatic}
\end{figure}

\begin{equation}
L_{int}=-{1\over 2}g^2 \phi'^2 \chi^2-h \bar{\psi}\psi \chi~.
\label{lagrangian}
\end{equation}
To avoid confusion, we drop the prime on $\phi$ remembering that $\phi < 0$
after inflation has ended. 
It should be noticed that the $\chi$ field has no bare mass,
its effective mass being determined by the field $\phi$ and the value of the coupling 
constant $g$ ($m_{\chi}=g\vert\phi\vert$).  

The production of $\chi$ particles commences as soon as $m_{\chi}$ begins changing
non-adiabatically \cite{FKL,FKL1}
\begin{equation}
|\dot{m_{\chi}}| \ggeq m_{\chi}^2~~~or~~~~|\dot{\phi}| \ggeq g\phi^2~.
\label{ad}
\end{equation}
The condition for particle production (\ref{ad}) is satisfied when 
\begin{equation}
\vert \phi\vert \lleq \vert \phi_{prod}\vert 
= \sqrt{\frac{\vert\dot{\phi}_{end}\vert}{g}}
=\sqrt{{V_{end}^{1/2} \over {\sqrt{3} g}}}~.
\label{phiprod}
\end{equation}
From (\ref{vend}) we find that $\phi_{prod}\ll M_p$ for $g\gg 10^{-9}$. 
The production time for $\chi$ particles can
be estimated to be 
\begin{equation}
\Delta t_{prod} \sim {\vert \phi\vert \over {|\dot{\phi}|}} 
\sim \frac{1}{\sqrt{g V_{end}^{1/2}}}~.
\label{tprod}
\end{equation}
The uncertainty relation provides an estimate for the momentum of $\chi$ particles
created non-adiabatically: $k_{prod} \simeq (\Delta t_{prod})^{-1}
\sim g^{1/2} V_{end}^{1/4}$. Proceeding as in \cite{kls97,FKL1} we can show that
the occupation number of $\chi$ particles jumps sharply from zero to 
\begin{equation}
n_k \simeq \exp(-{\pi k^2/{g V_{end}}^{1/2}})~,
\end{equation}
during the time interval $\Delta t_{prod}$.
The $\chi$-particle number density is estimated to be
\begin{equation}
n_{\chi}={1 \over {2\pi^3}} \int_0^{\infty}{k^2n_k} dk 
\simeq {(gV_{end}^{1/2})^{3/2} \over {8\pi^3}}~.
\label{nchi}
\end{equation}
Quanta of the $\chi$-field are created during the time interval $\Delta t_{prod}$
that the field $\phi$ spends in the vicinity of $\phi = 0$. Thereafter the mass 
of the $\chi$-particle begins to grow since $m_\chi = g\vert\phi(t)\vert$, and 
the energy density of particles of the $\chi$-field created in this manner is
given by
\begin{equation}
\rho_{\chi} = m_\chi n_{\chi}\left({a_{end} \over a}\right)^3 = 
{(gV^{1/2}_{end})^{3/2} \over 
{8\pi^3}} {{g \vert\phi(t)\vert}}\left({a_{end} \over a}\right)^3~.
\end{equation}
where the $({a_{end} / a})^3$ term accounts for the cosmological
dilution of the energy density with time.
As shown above, the process of $\chi$ particle-production 
takes place immediately after inflation has ended, provided $g \ggeq 10^{-9}$.
In what follows we will show that the $\chi$-field can rapidly decay into fermions. 
It is easy to show that if the quanta of the $\chi$-field were converted (thermalized)
into radiation instantaneously, the radiation energy density would become
\begin{equation}
\rho_r \simeq \rho_\chi \sim  {(gV^{1/2}_{end})^{3/2} \over {8\pi^3}} g \phi_{prod} \sim 
10^{-2} g^2 V_{end}~.
\label{irad}
\end{equation}
From equation (\ref{irad}) follows the important result
\begin{equation}
\left({\rho_\phi  \over \rho_r}\right)_{end}  \sim \left (\frac{10}{g}\right )^2~.
\label{irad1}
\end{equation}

\begin{figure}
\resizebox{3.0in}{!}{\includegraphics{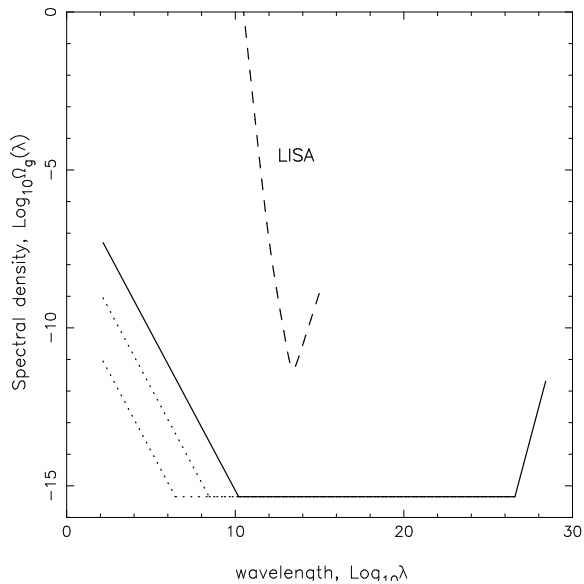}}
\caption{ The spectral energy density of the relic gravity wave background generated
by quintessential inflation on the brane followed by instant preheating.
The blue spectrum at short wavelengths is caused by the kinetic regime
which commences soon after brane inflation ends.
(The inflaton potential in (\ref{exp}) is assumed with $\alpha = 5$.)
At these wavelengths the solid line corresponds to $g = 4\times 10^{-3}$
and the
upper (lower) dashed line corresponds to $g = 0.03$ ($g = 0.3$), where $g$ is
the coupling parameter introduced in (\ref{lagrangian}). In all three cases
the gravity wave
background satisfies nucleosynthesis constraints. 
(The short-wavelength cutoff at $\lambda \simeq 10$ cm corresponds to the comoving
length scale marking the commencement of the kinetic regime. 
The spectrum of gravity waves with $\lambda \lleq 10$ cm will depend sensitively
upon the details of the preheating process and is not shown here.)
The chained line shows the expected sensitivity curve of the proposed Laser
Interferometer Space Antenna (LISA).
The present value of the Hubble parameter is assumed to be $H_0 = 70$ km/sec/Mpc.
} 
  \label{gwave}
  \end{figure}

Comparing (\ref{irad1}) with (\ref{endratio1}) we find that, in order for
relic gravity waves to respect the nucleosynthesis constraint, we should
have $g \ggeq 4\times 10^{-3}$; the corresponding gravity wave background is
shown in figure \ref{gwave}.
(The energy density created by instant preheating 
$\left (\rho_r/\rho_\phi\right ) \simeq (g/10)^2$ can clearly be 
much larger than the energy density
produced by quantum particle production, for which 
$\left (\rho_r/\rho_\phi\right ) \simeq 10^{-16}g_p$.)

The constraint
$g \ggeq 4\times 10^{-3}$, implies that the particle production time-scale
(\ref{tprod}) is much smaller than the Hubble time since
\beq
\frac{1}{\Delta t_{prod} H_{end}} \ggeq 300 \alpha^2, ~~\alpha \gg 1~.
\eeq
Thus the effects of expansion can safely be neglected 
during the very short time interval in which
`instant preheating' takes place. 
We also find, from equation (\ref{phiprod}), that $\vert \phi_{prod}\vert
/M_p \lleq 10^{-3}$
implying that particle production takes place in a very narrow band around
$\phi = 0$.
Figure \ref{nadiabatic} demonstrates
the violation of the adiabaticity condition (at the end of inflation) which is a
necessary prerequisite
for particle production to take place.
For the range of $g$
allowed by
the nucleosynthesis constraint, the particle production turns out to be almost instantaneous.

We now briefly address the issue of back-reaction of created $\chi$-particles
on the background. 
As mentioned above, brane effects persist briefly after inflation has 
ended. $\rho_{\phi} \propto 1/a$ immediately after inflation,
however, once the damping due to the $\rho^2$ term in
(\ref{friedman}) switches off, $\rho_{\phi}$ decreases faster,
settling to $\rho_{\phi} \propto 1/a^6$ at the commencement of
the kinetic regime. The transition period between the end of
inflation and the commencement of the kinetic regime
depends upon the value of $\alpha$, as reflected in (\ref{Hkin})
(for instance $a_{kin}/a_{end} \simeq 10$, for $\alpha =5$). The density of $\chi$ particles scales 
$\propto 1/a^3$ and modifies the field evolution equation
\begin{equation}
\ddot{\phi}(t)+3H\dot{\phi}(t)+V_{,\phi}+g^2 \phi \langle \chi^2 \rangle =0
\label{eqbreaction}
\end{equation}    
where $\langle \chi^2 \rangle$ can estimated according to the prescription given in 
Refs. \cite{FKL} and \cite{FKL1}
\begin{equation}
 \langle  \chi^2\rangle  ={{(g V^{1/2}_{end})^{3/2}} \over {8\pi^3 \sqrt{3} g \phi}} ({a_{end} \over a})^3~.
\end{equation}
\hskip-0.9cm
\begin{figure}
\resizebox{3.8in}{!}{\includegraphics{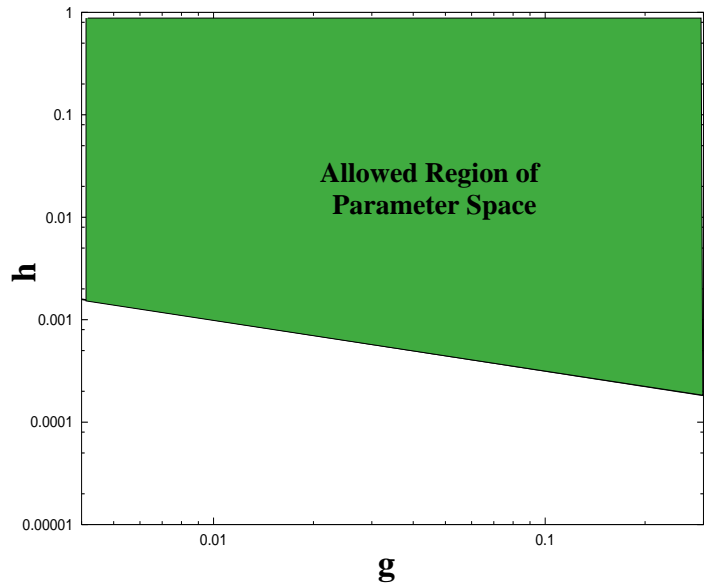}}
\caption{Parameter space permitted by nucleosynthesis 
constraint and the constraint imposed by the back reaction in models of brane inflation with instant preheating. The upper limits on
the parameters $h$ is 
dictated by the validity of perturbation theory for the interaction described
by the Lagrangian (\ref{lagrangian}). Clearly there is a large
region in parameter space for which the
{\it instant preheating} mechanism can be used to construct realistic models of
quintessential inflation. 
}
\label{parameter}
\end{figure} 

The density of non-relativistic $\chi$ particles $\rho_{\chi}$ scales as 
$1/a^3$. In contrast, the potential term, as well as the
dissipative terms  in the field equation scale slower than $\rho_\chi$ 
for most of the evolution before the commencement of the kinetic regime. 
As a result, for any generic value of the coupling $g \lleq 0.3$, the back-reaction 
of $\chi$ particles in the evolution equation is negligible during the time
scale $\sim H_{kin}^{-1}$. 
It is important to note that, although $\rho_\chi/\rho_\phi$ is a small
quantity to begin with, $\rho_\chi$ (or the decay products of the $\chi$ field)
will decrease as $1/a^n(t)$, $n \leq 4$.
In contrast, since the scalar field is rolling down an exponentially steep
potential, its energy density decreases much faster $\rho_\phi \propto 1/a^6(t)$
during the kinetic regime.
Therefore either $\rho_\chi$ or its decay products 
will eventually come to dominate the density
of the universe. 

We now turn to the matter of reheating which occurs through the
decay of $\chi$ particles to fermions, as a consequence of the interaction term in the
Lagrangian (\ref{lagrangian}). 
The decay rate of $\chi$ particles is given by
$\Gamma_{\bar{\psi}\psi}=h^2m_{\chi}/8\pi$, where $m_{\chi} = g\vert\phi\vert$. 
Clearly the decay rate is faster for larger values of $\vert\phi\vert$.
For $\Gamma_{\bar{\psi}\psi} >H_{kin}$, the decay process will
be completed within the time that back-reaction effects (of $\chi$ particles)
remain small. Using (\ref{Hkin}) this requirement translates into
\begin{equation}
h^2 > {{8\pi \alpha} \over {\sqrt{3} g\phi}} {V_{end}^{1/2} \over M_p}F(\alpha)~.
\label{Gamma}
\end{equation}
For reheating to be completed by $\phi/M_p \lleq 1$, 
we find  from  equations (\ref{endratio1}) and 
 (\ref{Gamma})  that $ h \ggeq 10^{-4} g^{-1/2}$ for $\alpha \simeq 5$.
In figure \ref{parameter} we have shown the region $ h \ggeq 10^{-4} g^{-1/2}$,
$g \ggeq 4 \times 10^{-3}$, for which (i) reheating is  rapid and
(ii) the relic gravity background in non-oscillatory braneworld models of 
quintessential inflation
is consistent with nucleo-synthesis constraints. 
As discussed in \cite{FKL} this method of reheating can give rise to 
super-heavy fermions
with $m_\psi \sim 10^{16} - 10^{17}$ GeV, the subsequent decay of these particles
completes the reheating process.

The method discussed here can also be extended to long lived quanta of the
$\chi$ field with bare mass $\tilde{m_\chi}$. In this case it can be shown that
the probability of creation of such particles will be suppressed by the factor
(see also \cite{FKL})
\beq
\exp{\left \lbrack -\frac{\pi {\tilde m_\chi}^2}{g |{\dot\phi}|}\right \rbrack}
= \exp{\left \lbrack -1.7~ 10^7 \left (\frac{\pi \alpha^2}{g}\right )
\left (\frac{{\tilde m_\chi}}{M_p}\right )^2
\right \rbrack}~,
\label{suppressed}
\eeq
where we have used (\ref{velo}) and (\ref{vend}). 
From (\ref{suppressed}) we find that the creation of long lived quasi-stable
particles of mass $\tilde{m_\chi} \simeq 2 \times 10^{15}$ GeV is strongly 
suppressed by the factor $8 \times 10^{-17}$ if $\alpha = 5$ and $g \sim  1$.
Since the density of quasi-stable particles decreases as $1/a^3(t)$, 
this mechanism
allows us to create a small number of super-heavy WIMPs after braneworld inflation
has ended and
whose role becomes important at late times, when they can be candidates for
dark matter or, if they decay,
be capable of producing cosmic rays more energetic than the GZK limit \cite{gzk}
(see \cite{FKL,ckr98,chung98} for related extensive discussions of this issue).

The analysis presented here is also likely to qualitatively
hold for a class of inverse square potentials.
We conclude that instant preheating in the context of brane inflation is very 
efficient and is capable of producing successful models of non-oscillatory
quintessential inflation on the brane.

\section{Late-Time Behaviour}

\bigskip
\begin{figure}
\vspace{0.5in}
\resizebox{3.5in}{!}{\includegraphics{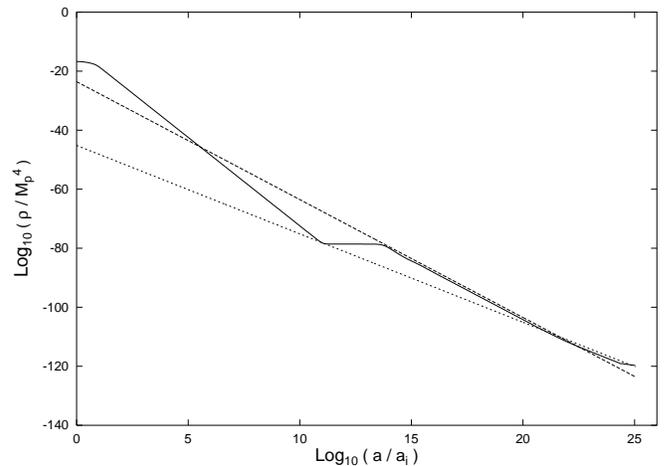}}
\caption{The post-inflationary evolution of the scalar field energy density
(solid line),
radiation (dashed line) and cold dark matter (dotted line) is shown as a
function of the scale factor for the quintessential inflation model described by
(\ref{cosine})
with $V_0^{1/4} \simeq 10^{-30} M_p$,
$\tilde{\alpha}=50$ and $p=0.1$ ($\alpha=p \tilde{\alpha}=5$).
After brane effects have ended, the field energy density $\rho_{\phi}$ enters the
kinetic regime and soon drops below the radiation density.
After a brief interval during which $<w_\phi> \simeq -1$, the scalar field begins
to track first radiation and then matter.
At very late times (present epoch) the scalar field plays the role of quintessence
and makes the universe accelerate.
The evolution of the energy density
is shown from the end of inflation until the present epoch.
}
\label{den}
\end{figure}

It is well known that, during the post-brane epoch, 
a scaling solution is an attractor for the
exponential potential (see \cite{fj97b}
and references therein). 
After the attractor is reached, the field energy density
tracks the background in such a way that $\rho_{\phi}/\rho_r$ remains constant. 
Compliance with nucleosynthesis constraints requires $\Omega_{\phi} \lleq 0.2$
during the radiative regime, which implies $\alpha \ggeq 5$. 
However, for a purely exponential potential
$\rho_{\phi}$ remains subdominant forever and the scalar field never 
becomes quintessence. Quintessential inflation can be achieved 
for a potential which interpolates between an exponential and a flat or 
rapidly oscillating form. An example of the latter is provided by \cite{sw00}
\begin{equation}
V(\phi)=V_0\left[\cosh ({\tilde \alpha} \phi/M_p)-1\right]^p,~~~~0 < ~p < ~1/2 
\label{cosine}
\end{equation}
which has asymptotic forms
\begin{equation}
V(\phi)={V_0 \over 2^p} e^{{\tilde\alpha}  p \phi/M_p},~~~\tilde\alpha  \vert\phi\vert /M_p~>>1~,
\label{exppot}
\end{equation}
and
\begin{equation}
V(\phi)={V_0 \over 2^p}\left({\tilde\alpha} \phi \over M_p \right)^{2p}
~~~~~~\tilde\alpha  |\phi|/M_p ~<< 1~,
\label{ppot}
\end{equation}
where ${\tilde \alpha}p = \alpha$.
The power law behaviour of (\ref{cosine}) near the origin leads  to oscillations
of $\phi$ when it  approaches small values.
Rewriting $\phi$ in terms of $\phi'$ as, 
$\phi=\phi'+\phi_{end}$, reproduces the correct asymptotes (\ref{texp}) 
for large values of
$\phi'$ as well as the power-law behavior (\ref{ppot}) for $\phi' \to -\phi_{end}$.
The time-averaged equation of state of the scalar field during oscillations is
given by \cite{sw00} 
\begin{equation}
\langle w_{\phi} \rangle ={{p-1} \over{p+1}}~,
\end{equation}
as a result, the scalar-field energy density and the scale factor display
the following behavior
$$\rho_{\phi} \propto  a^{-3(1+\langle w_{\phi} \rangle)},~~~~~~~~
a \propto   t^{{ 2 \over 3}(1 + \langle w_{\phi} \rangle )^{-1}}~. $$

Clearly, the scalar field behaves like dark energy for $p<1/2$. By 
adjusting parameters in this model one can  ensure that scalar field 
oscillations occur at late times so that $\Omega_{\phi}$ reaches 0.7 today. 
We have evolved the field
equations numerically and our results (for a particular parameter choice)
are shown in
figure \ref{den}.    

Other forms of the potential which will give rise to quintessential inflation
can be found in \cite{sahni02a}.
\bigskip
\begin{figure}
\vspace{0.5in}
\resizebox{3.5in}{!}{\includegraphics{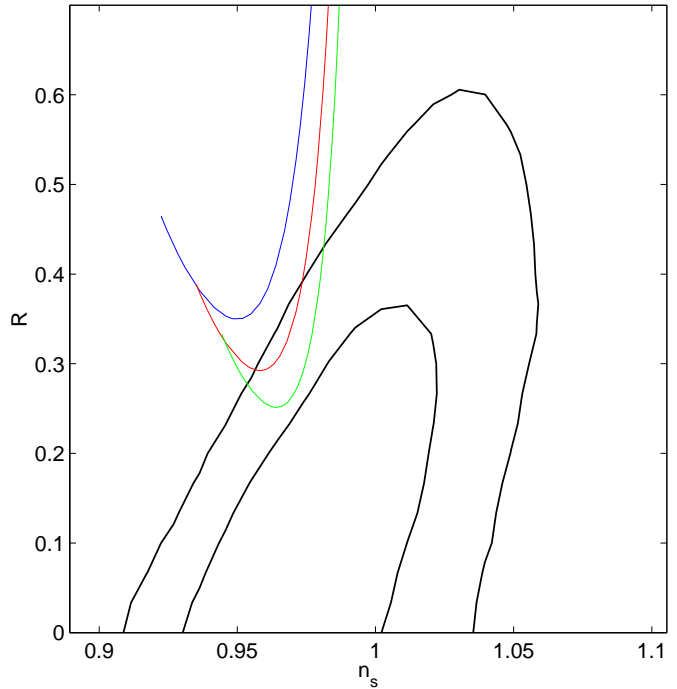}}
\caption{Plot of $R \equiv 16 A_T^2/A_S^2$ 
versus the spectral index $n_S$ for inflation driven by the 
exponential potential (\ref{exp}) in a Gauss-Bonnet (GB) braneworld 
for ${\cal N}=50,60,70$ e-foldings (from top to bottom). 
The $1\sigma$ and $2\sigma$ observational contours which follow from the 
WMAP+SDSS results are shown for comparison \cite{ssr}. 
The $R(n_S)$ curves exhibit a minimum in the
intermediate energy region between the GB (extreme right) and 
RS (extreme left) regimes. It is clear that for a larger
value of ${\cal N}$ (the theoretical curves are outside
 the $2 \sigma$ contour bond for $N=50$, however the points begin to move inside the
 $2\sigma$ bound as ${\cal N}$ becomes larger than $55$) more points in this region lie within the $2\sigma$ 
observational bound illustrating the fact that steep inflation in the GB model
is in agreement with observations (see \cite{ssr} for more details).
}
\label{exponential.ps}
\end{figure}

\section{Conclusions}

In this paper we have successfully applied the instant preheating mechanism 
discussed in \cite{FKL,FKL1} to braneworld inflation. 
Braneworld cosmology has the attractive property of allowing inflation
to occur even for steep potentials thus greatly expanding the class of potentials
which give rise to inflation. We have demonstrated that instant preheating
works remarkably well for non-oscillating potentials such as an exponential.
In such models the instant preheating mechanism 
results in
a much higher energy density for radiation
than the mechanism based on gravitational particle production \cite{ford,spokoiny}.

One of the main results of this paper is
that a single scalar field on the brane can successfully 
describe inflation at early epochs and quintessence at late times. 
Braneworld models of quintessential inflation followed by instant preheating 
do not over-produce gravity waves and are therefore
entirely consistent with the supernova data 
on the one hand and 
nucleosynthesis constraints on the other. 
Although the amplitude of the gravity wave background in quintessential inflation
is below
the projected sensitivity of LISA, the `blue spectrum' of gravity waves
with wavelengths $\lleq 10^{10}$ cm, makes it likely that that they could lie
within the sensitivity range of
future space-based gravity wave detectors
(see for example \cite{seto01}).

We would like to end by mentioning that the main features of instant preheating in
braneworld models are sufficiently robust and are likely to carry over to other 
inflationary models in which enhanced damping is an important feature of early scalar
field dynamics. 
In this paper we have laid stress on the `heavy damping' experienced by a scalar
field during the RS phase in order to construct a successful scenario of
quintessential-inflation on the brane.
It should be emphasised however that the RS phase also occurs in braneworld models
which have a Gauss-Bonnet (GB) term in the bulk; see equations 
(\ref{action}) - (\ref{gr}). Indeed it is to this class of models to which one must
turn for a scenario, which not only gives rise to quintessential-inflation,
but also satisfies all other observational constraints -- particularly those 
on the spectral index and the scalar to tensor ratio placed by WMAP+SDSS
\cite{spergel03,tegmark03}. It is well known that steep inflation in the
RS scenario comes into conflict with observations \cite{suji04} while 
GB brane world can rescue these models \cite{ssr}.
Indeed, it appears that steep inflation in GB models which commences
at an intermediate energy scale between RS and GB regimes
-- can be easily reconciled with observations \cite{ssr}.
Figure \ref{exponential.ps}
summarizes this result by showing the values of the scalar to tensor
ratio $R$ and the scalar spectral index $n_S$ in the GB inflationary model 
; for details the reader is referred to 
\cite{ssr}. It is interesting that the $R(n_S)$ curve shows a minimum 
for steep inflation which begins at an
intermediate energy scale in GB inflation and which agrees with
observations. 
(The upper right branch of the curves
corresponds to steep inflation which commenced deep in the GB regime; 
the left branch corresponds to the RS regime.)
We conclude that
a successful scenario of quintessential inflation on the 
Gauss-Bonnet braneworld has been
constructed which agrees
with CMB+LSS observations and also generates an interesting blue spectrum
for gravity waves on small scales. 

\bigskip

\section{Acknowledgments} 
We acknowledge useful discussions with Parampreet Singh and Yuri Shtanov and 
I. P. Neupane. We are indebted to Shinji Tsujikawa
for supporting the likelihood analysis and for making critical comments.

\end{document}